\documentclass[11pt, a4paper, pdftex, openany, dvipsnames]{article}

\usepackage{makecell}
\usepackage{soul}
\usepackage{forest}
\usepackage{tikz}
\usetikzlibrary{arrows.meta, positioning, calc, shapes.multipart}

\usepackage{tikz}
\usetikzlibrary{calc,quotes,angles,intersections}
\usepackage{pgfplots}

\usepackage{booktabs}
\usepackage{array}
\usepackage{amsmath}
\usepackage{algorithm}
\usepackage{algpseudocode}  

\usepackage{algpseudocode}
\usepackage{braket}

\usepackage[breakable]{tcolorbox} 
\usepackage[margin=1in]{geometry}  

\let\originalparagraph\paragraph
    \renewcommand{\paragraph}[2][.]{\originalparagraph{#2#1}}
\usepackage{pgfplots}
\usepackage{amsmath}

\pgfplotsset{compat=1.17}
\usepackage{natbib}
\setcitestyle{citesep={,}}
\usepackage{url}
\usepackage{appendix}
\usepackage{amsfonts}
\usepackage{amsmath,amssymb}
\usepackage{bm}
\usepackage{mathrsfs}
\usepackage{mathtools}
\usepackage{graphics}
\usepackage{graphicx}
\usepackage{color}
\usepackage{theorem}
\usepackage{setspace}
\usepackage{fullpage}
\usepackage{longtable}
\usepackage[T1]{fontenc}
\usepackage{makecell}
\usepackage{newpxtext,newpxmath}
\usepackage{caption}
\usepackage{xcolor}
\usepackage{hyperref}
\usepackage{comment}

\hypersetup{
    colorlinks = true,
    linkcolor=blue,
    citecolor=red,
    urlcolor=cyan
    }

\newcommand{\bbb}[1]{\bar{\boldsymbol{\mathrm{#1}}}}

\newcommand{\card}[1]{|#1|}
\newcommand{\ekk}[0]{\mathcal{K}}

\usepackage[margin=1in]{geometry}

\newcommand{\fixmeSimon}[1]{\textbf{\emph{\textcolor{blue}{/* TBD Simon: #1 */}}}}
\newcommand{\fixmeNormann}[1]{\textbf{\emph{\textcolor{red}{/* TBD Normann: #1 */}}}}

\newcommand{\argmin}[1]{\underset{#1}{\text{argmin}} }

\newcommand{\bb}[1]{\boldsymbol{\bold{#1}}}

\newcommand{\eval}[2]{|_{#1=#2}}

\newcommand\norm[2]{\left\lVert #1 \right\rVert_{#2}}

\renewcommand{\card}[1]{|#1|}

\makeatletter
\renewcommand*{\@fnsymbol}[1]{\ifcase#1\or*\else\@arabic{\numexpr#1-1\relax}\fi}
\makeatother

\title{Scalable Global Solution Techniques for High-Dimensional Models in Dynare}

\author{\small
\makebox[\linewidth][c]{%
\begin{tabular}{cccc}
\makecell{
    Aryan Eftekhari \\ 
    Institute of Computing \\ 
    USI Lugano \\ 
    Switzerland \\
    \texttt{aryan.eftekhari@usi.ch}
}
&
\makecell{
    Michel Juillard \\ 
    Banque de France \\ 
    \\
    France\\ 
    \texttt{michel.juillard@mjui.fr}
}
&
\makecell{
    Normann Rion \\ 
    CY Cergy Paris Universit\'e \\ 
    CEPREMAP\\ 
    France\\ 
    \texttt{normann@dynare.org}
}
&
\makecell{
    Simon Scheidegger\footnote{We thank Lorenzo Bretscher, Johannes Brumm, Aurélien Eyquem, Galo Nu\~no, Tam\'as Simon, D\'aniel Sali, Gauthier Vermandel, and the participants of the 
    ``Celebrating Michel Juillard's Career and 30 Years of Dynare'' conference for their valuable comments. This work is supported by a grant
    from the Swiss National Science Foundation under project ID
    “New Methods for Asset Pricing with Frictions.”} \\ 
    Department of Economics \\ 
    University of Lausanne  \\ 
    Switzerland \\
    \texttt{simon.scheidegger@unil.ch}
}
\end{tabular}%
}
}

\date{\today}

\begin{document}
\let\cleardoublepage\clearpage

	\begin{singlespace}

\maketitle
\vspace{-.4in}

\abstract{For over three decades, Dynare has been a cornerstone of dynamic stochastic modeling in economics, relying primarily on perturbation-based local solution methods. However, these techniques often falter in high-dimensional, non-linear models that demand more comprehensive approaches.
This paper demonstrates that global solutions of economic models with substantial heterogeneity and frictions can be computed accurately and swiftly by augmenting Dynare with adaptive sparse grids (SGs) and high-dimensional model representation (HDMR). 
SGs mitigate the curse of dimensionality, as the number of grid points grows significantly slower than in traditional tensor-product Cartesian grids. Additionally, adaptivity focuses grid refinement on regions with steep gradients or non-differentiabilities, enhancing computational efficiency. Complementing SGs, HDMR tackles large state spaces by approximating policy functions with a hierarchical expansion of low-dimensional terms. 
Using a time iteration algorithm, we benchmark our approach on an international real business cycle model.
Our results show that both SGs and HDMR alleviate the curse of dimensionality, enabling accurate solutions for at least 100-dimensional models on standard hardware in relatively short times. This advancement extends Dynare’s capabilities beyond perturbation approaches, establishing a versatile platform for sophisticated non-linear models and paving the way for integrating the most recent global solution methods, such as those from machine learning.}

\medskip

\noindent \textit{JEL classification}: C63, E30, F44.

\medskip
\noindent \textit{Keywords}: Adaptive Sparse Grids, High-dimensional Model Representation, Global Solution Methods, International Real Business Cycles, Time Iteration.
\end{singlespace}

\clearpage

\setcounter{footnote}{1}

	\begin{singlespace}

\section{Introduction} 
\label{secintroduction}

\paragraph{Motivation} Over the past three decades, Dynare~\citep{juillard1994dynare,COLLARD2001979,Adjemianetal2024} has become an indispensable tool for researchers and policymakers in macroeconomics and beyond. Its guiding principle, \textit{``write your model almost as you would on paper, and Dynare will take care of the rest!''}, offers a user-friendly approach that has led to widespread adoption in academic and policy circles alike.

Dynare’s intuitive framework and local solution methods have transformed the specification and analysis of dynamic stochastic models in economics. Yet, these methods struggle with problems involving substantial heterogeneity and non-linearities that demand global solutions,\footnote{We adopt the nomenclature from \citet{brummscheidegger_2017}, referring to a “global solution” as one computed using equilibrium conditions at numerous points throughout the state space of a dynamic model, as opposed to a “local solution” which relies on a local approximation around the model's steady state, as achieved through perturbation methods.} such as those with financial frictions, the zero lower bound, rare disasters, and models with many agents (e.g.,~\citealp{kruegerkubler_2004,brumm2015collateral,fernandez-villaverde_2015,azinovic2023economics}), or tipping points in climate-economic models (e.g.,~\citealp{doi:10.1086/701890,kotlikoffetal_2020_WP}). While perturbation methods are confined to a neighborhood of the steady state, \textit{naive} global solution methods, such as basic Cartesian grid-based approaches requiring \(M^d\) points for \(M\) points per dimension, face exponential computational costs as dimensionality $d$ grows. This challenge, driven, for instance, by the heterogeneity in a model, is commonly known as the \textit{curse of dimensionality}~\citep{bellman_1961}.

\paragraph{Goal and Contribution of this Article}\vspace{-1.25\baselineskip}
Recognizing the limitations of local solution methods, this paper demonstrates that global solutions for economic models with substantial heterogeneity and frictions can be computed both accurately and fast by extending the Dynare environment—specifically, its \texttt{.mod}-files—to accommodate contemporary global solution techniques. We illustrate these enhancements by integrating two widely used and well-studied methods—(adaptive) sparse grids (SGs\footnote{We use the term SGs interchangeably for both ``regular'' sparse grids with piecewise linear basis functions 
and ``adaptive'' sparse grids with linear basis functions (cf. Section~\ref{sec:3.1}).
}; e.g., \citealp{brummscheidegger_2017}) and high-dimensional model representation (HDMR; e.g., \citealp{eftekharietal_2017})—and by showcasing their generic applicability and scalability through the solution of an international real business cycle (IRBC) model \citep{DenHaan2011175,kollmannetal_2011} via the time iteration algorithm~\citep{coleman1990}.\footnote{Two main challenges arise when applying time iteration to large-scale dynamic stochastic models: 
(i) each iteration step requires the global approximation of a high-dimensional, multivariate function, and (ii) each point in the chosen approximation scheme entails solving a system of non-linear equations. 
Together, these issues can significantly prolong the time to solution, underscoring the need for an efficient and highly scalable implementation of the algorithm.} The IRBC model is an ideal test case because its dimensionality grows linearly with the number of countries, making it well-suited for testing high-dimensional global solution methods.

\paragraph{General Problem Formulation and Global Solution Methods}\vspace{-1.25\baselineskip} 
This article examines dynamic economic models frequently characterized by \emph{recursive equilibria} \citep{ljungqvist2004recursive}. In these models, a (potentially high-dimensional) \emph{state variable} $\bb{x} \in X \subset \mathbb{R}^d$ represents the current state of the economy, with $d$ indicating the dimensionality of the state space. The evolution of the model is governed by a time-invariant \emph{equilibrium function} $f : X \to Y \subset \mathbb{R}^m$, which is determined by solving a functional equation of the form
\begin{equation}
    \mathcal{E}(f)=\mathbf{0},
\label{eq:DynamicEq}
\end{equation}
where $\mathcal{E}$ may, for example, denote a Bellman equation in a discrete-time framework or the Hamilton-Jacobi-Bellman (HJB) equation in continuous time. Alternatively, $\mathcal{E}$ might capture the first-order equilibrium conditions in a discrete-time setting, with $f$ representing a (possibly multi-dimensional) policy function, which is the primary focus of this article.

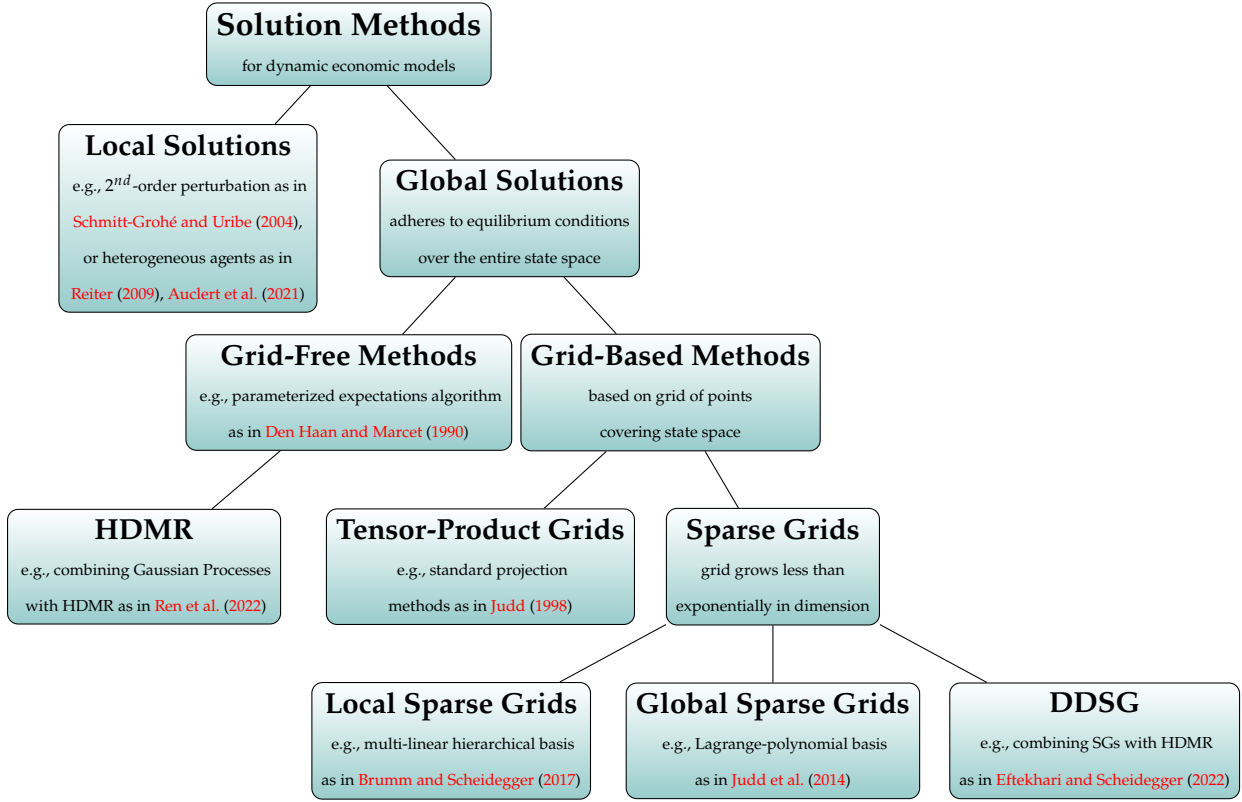
\begin{figure}[t!]
\centering{\title{Overview of Numerical Solution Methods for Dynamic Models.\vfill \break}}
\bigskip
\centering		
\begin{tikzpicture}[sibling distance=11em, level distance=6em,
  every node/.style = {shape=rectangle, rounded corners,
    draw, align=center,
    top color=white, bottom color=teal!40}]
  \node {{\large\bf{Solution Methods}} \\ \tiny{for dynamic economic models}}
    child  { node {{\bf{Local Solutions}}\\ \tiny{ e.g.,\ $2^{nd}$-order perturbation as in} \\ \tiny{\cite{schmitt2004solving},}\\ \tiny{or heterogeneous agents as in} \\ \tiny{\cite{reiter_2009,https://doi.org/10.3982/ECTA17434}} } }
    child { node {{\bf{Global Solutions}}\\ \tiny{adheres to equilibrium conditions}\\ \tiny{over the entire state space} }
      child { node {{\bf{Grid-Free Methods}} \\ \tiny{ e.g.,\ parameterized expectations algorithm} \\ \tiny{as in \cite{den1990solving}}}
        child [xshift=-7em] { node {{\bf{HDMR}} \\ \tiny{ e.g.,\ combining Gaussian Processes} \\ \tiny{ with HDMR as in~\cite{REN2022108220}}} }
      }
      child { node {{\bf{Grid-Based Methods}}\\ \tiny{based on grid of points}\\ \tiny{covering state space}} 
        child [xshift=-1em] { node {{\bf{Tensor-Product Grids}} \\ \tiny{ e.g.,\ standard projection } \\ \tiny{methods as in \cite{judd_1998}}}}
        child [xshift=-2em]{ node {{\bf{Sparse Grids}}\\ \tiny{grid grows less than}\\ \tiny{exponentially in dimension}}
          child { node {{\bf{Local Sparse Grids}} \\ \tiny{ e.g.,\ multi-linear hierarchical basis } \\ \tiny{as in \cite{brummscheidegger_2017}}}}
          child { node {{\bf{Global Sparse Grids}} \\ \tiny{ e.g.,\ Lagrange-polynomial basis } \\ \tiny{as in \cite{juddetal_2014}}}}
          child { node {{\bf{DDSG}} \\ \tiny{ e.g.,\ combining SGs with HDMR} \\ \tiny{as in \cite{doi:10.1137/21M1392231}}}
        }
      }
    }
  };
\end{tikzpicture}
\caption{Taxonomy of solution methods for dynamic economic models, classified into several categories. Note that HDMR can be utilized in both grid-based and grid-free contexts. However, in this article, we focus exclusively on grid-based approaches, as the combination of SGs with HDMR (denoted as \emph{dimension-decomposed sparse grids} (DDSG); cf. Section~\ref{sec:3.3}) offers significant numerical advantages.
 \label{fig:LitRev}}
\end{figure}

In each of these settings, the \emph{curse of dimensionality} becomes an issue when the state space $X$ is moderately high-dimensional and a global solution is required, particularly in the presence of significant non-linearities. In contrast to local solutions, which depend on equilibrium conditions and their derivatives at a specific point, global approaches demand that these conditions hold throughout the entire state space. Consequently, researchers have to turn to grid-based and grid-free approximation methods that mitigate the curse of dimensionality. 
Figure~\ref{fig:LitRev} provides a structured, albeit non-exhaustive, taxonomy of solution methods used in dynamic economic models, adapted and expanded from~\cite{Brumm2022}.
%

Grid-free methods have been proposed, notably by \citet{den1990solving}, and have recently gained traction through advancements in machine learning.\footnote{See, for instance,~\citealp{duffy_mcnelis_2001,norets_2012, rennerscheidegger_2017,MALIAR202176,azinovic2022,FV2023,Han2021,Friedl2023,gaegauf2023comprehensive,nuno2024monetary,https://doi.org/10.3982/qe1403,10.1093/rfs/hhae043,kase2022,kubler2025,Payne2024}, and~\citealp{chen2025dynamic}. For a recent review of deep learning applications in economics, refer to~\cite{NBERw33117}.} Unlike many contemporary machine learning approaches, which often require extensive ad-hoc hyperparameter tuning, SGs benefit from well-understood convergence properties. SGs enable researchers to scale up models that are numerically formulated on a grid to higher dimensions without necessitating a complete overhaul of the solution technique. Additionally, they can be relatively easily parallelized, potentially reducing computation time significantly~\citep{brumm201512}. Thus, SGs provide a straightforward means to augment current modeling and solution frameworks, thereby broadening the spectrum of addressable research questions. We have thus opted to initially employ SGs (in combination with HDRM) to extend Dynare beyond perturbation-based methods before transitioning to more recent, albeit currently less stable, global solution techniques.

\begin{figure}[t!]
    \centering
    \includegraphics[width=0.99\textwidth]{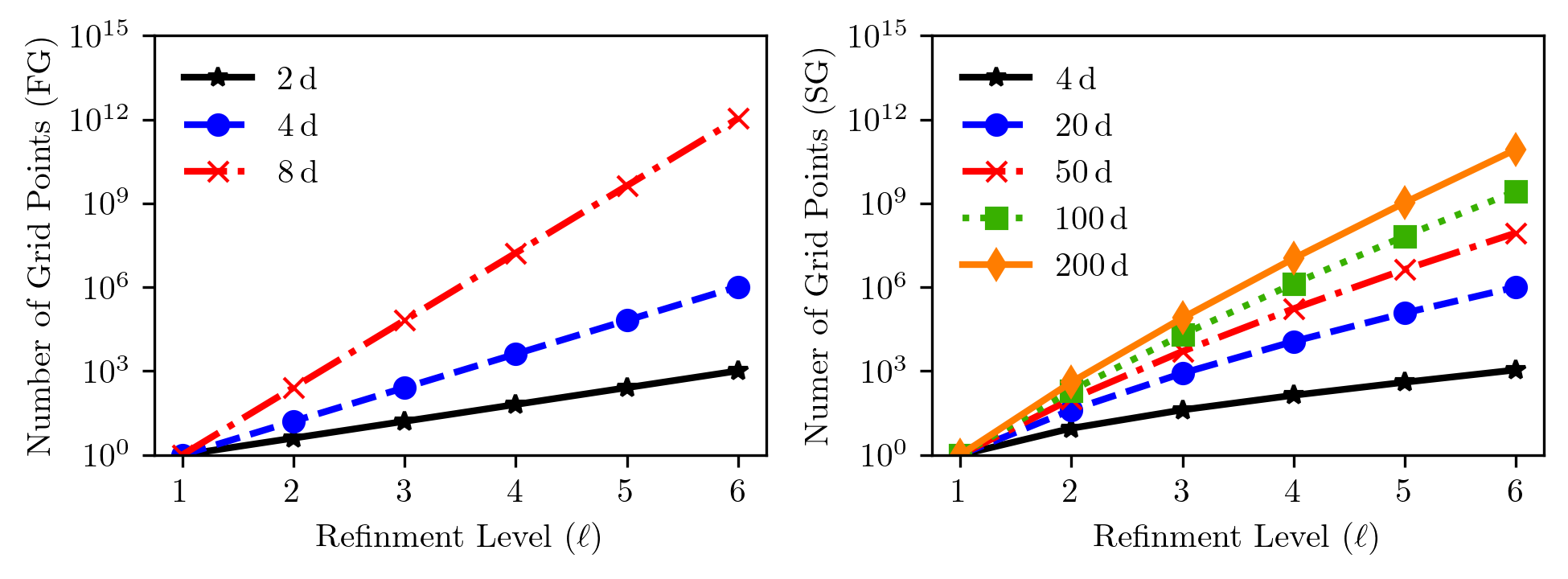}
    \caption{Left Panel: Number of grid points in a full Cartesian grid (FG) of dimensionality \(d\) with increasing grid resolution, denoted by the refinement level \(\ell\). In SGs, \(\ell\) dictates the grid's resolution, producing approximately \(M \propto 2^\ell\) points per dimension, where a higher \(\ell\) resolves finer features as the spacing between points shrinks to \(h \propto 2^{-\ell}\), resulting in a total of \(M \propto 2^{\ell \cdot d}\) grid points across all dimensions. Right Panel: Number of grid points in an SG of increasing dimensionality. Notably, even a 200-dimensional SG with resolution \(\ell=6\) has orders of magnitude fewer points than an 8-dimensional FG at the same resolution level (cf. Section~\ref{sec:3.1} below).
%
    }
    \label{fig:1}
\end{figure}

\paragraph{Adaptive Sparse Grids}\vspace{-1.25\baselineskip} 
SG methods offer a highly efficient and structured approach to the computational challenges of high-dimensional state spaces in dynamic economic models, that is, to approximate non-linear, high-dimensional (policy) functions. By extending univariate interpolation formulas to the multivariate case through linear combinations of tensor products, as detailed in various techniques like the Smolyak algorithm, classical SG methods, combination techniques, and dimension- or spatially-adaptive methods (e.g.,~\citealp{Bungartz.Griebel:2004}, and references therein), SGs alleviate the curse of dimensionality. Specifically, SG methods reduce the exponential growth of grid points with increasing dimensionality \(d\), from \(\mathcal{O}(M^d)\) to \(\mathcal{O}(M \cdot (\log M)^{d-1})\), while maintaining nearly the same accuracy for sufficiently smooth functions (see Figure~\ref{fig:1}). This approach cuts grid points by several orders of magnitude compared to full Cartesian tensor-product grids, significantly mitigating the computational burden while marginally increasing interpolation errors. The primary distinction among SG techniques lies in their use of either local or global basis functions, which we categorize as local sparse grids and global sparse grids, respectively. Local SGs (LSG), utilizing hierarchical, multi-linear basis functions, are particularly effective for non-smooth functions exhibiting localized irregularities like kinks. On the other hand, Global SGs (GSG), often based on Lagrange characteristic polynomials (commonly known as the Smolyak method in economics), excel at approximating smooth functions (cf. Figure~\ref{fig:LitRev}). In this article, we focus strictly on LSGs; for GSGs, please refer to the review by~\cite{Brumm2022}.

\paragraph{The Role of HDMR: When Sparse Grids Are Not Sufficient}\vspace{-1.25\baselineskip}
%

SGs are highly effective for approximating functions in moderately high-dimensional spaces (e.g., when \(d < 20\)). However, their computational cost can become prohibitive as either the dimensionality or the complexity (i.e., the degree of non-linearity) of the underlying functions increases.
For example, in highly non-linear dynamic economic models with many state variables, a high \emph{a priori} grid resolution (i.e., refinement \(\ell\)) produces an explosive number of grid points (see Figure~\ref{fig:1}). Moreover, numerical operations on SG data structures may become prohibitively expensive in high-dimensional settings~\citep{muraracsu2012fastsg}. HDMR addresses this limitation by decomposing the target function into a sum of lower-dimensional components, each of which can be approximated very efficiently (see, e.g.,~\citealp{Ma:2010:AHS:1751671.1751796,yangetal_2012}, and references therein). Evaluating the combined sum of these component functions is considerably less computationally demanding than evaluating a full SG in high-dimensional settings.
When HDMR is combined with SGs, the resulting approach is often called \emph{dimension-decomposed sparse grids} (DDSG; cf. Figure~\ref{fig:LitRev}). This integration substantially reduces computational load and memory requirements while preserving approximation accuracy. As a result, DDSG is particularly well-suited for high-dimensional economic models characterized by significant heterogeneity and complex non-linearities, provided that a (possibly latent) additively separable structure exists and can be detected computationally.




\paragraph{Preview of Results}\vspace{-1.25\baselineskip}
We demonstrate that the SG approach for computing global solutions scales up to at least 16-dimensional IRBC models\footnote{Recall that while we focus on an IRBC model here, the approach is broadly applicable to high-dimensional dynamic stochastic models. The IRBC model simply serves as a convenient test bed, because one can easily adjust the number of countries (and thus the number of states) to control the computational complexity.} and exhibits subexponential increases in runtime. 
Even when targeting a high level of accuracy for global solutions, our computation times remain relatively short on a standard laptop: 8-dimensional models are solved in seconds, while 16-dimensional ones require only about half an hour.
Moreover, initializing the time iteration algorithm with Dynare's perturbation solution—rather than a naive guess—reduces convergence time by up to 20 times, highlighting the practical viability of the SG method for high-dimensional models.
In addition, our DDSG approach, which builds upon standard SG methods, can provide additional efficiency gains, as demonstrated in the context of the IRBC model with its latent additively separable structure.
Although the DDSG method incurs a modest overhead in low-dimensional settings, this cost diminishes rapidly with increasing dimensions; runtimes break even by eight dimensions, and at 16 dimensions, a single DDSG timestep is more than 13 times faster than its SG counterpart. These efficiency gains, coupled with an up to 80 times speedup achieved by employing Dynare's initialization within the DDSG framework, make our approach a powerful tool for tackling complex, high-dimensional economic models in which latent additively separable structures are present. 
As we demonstrate in our numerical experiments, the DDSG approach also exhibits subexponential increases in runtime, thereby enabling the computation of global solutions for models with at least 100 dimensions on a standard laptop within hours, whereas models with fewer than 20 dimensions exhibit runtimes of only seconds to minutes.


\paragraph{Organization of the Article}\vspace{-1.25\baselineskip} 

The remainder of this paper is structured as follows. 
Section~\ref{sec:literature} offers a brief overview of related literature. 
Section~\ref{Sec:dynamicModels} then formally characterizes the models we aim to solve via a canonical time iteration algorithm. 
Section~\ref{sec:approximator} reviews the mathematical underpinnings of SGs and HDMR, which serve to approximate and interpolate policy functions within the time iteration algorithm. 
Section~\ref{sec:syntax} details the required modifications to Dynare’s \texttt{.mod} files to accommodate the global solution methods proposed. 
Section~\ref{sec:results} then discusses illustrative performance results, 
and finally, Section~\ref{sec:conclusion} concludes. 
Furthermore, we provide supplementary code examples demonstrating our developments, available at~\url{https://nvls.co/Dynare/GlobalMethods/SparseGrids}.
\section{Related Literature}
\label{sec:literature}

Building on \citealp{juillard1994dynare,COLLARD2001979} and \citealp{Adjemianetal2024}, this article aims to extend the Dynare framework—currently relying on various local perturbation methods—by incorporating two particular types of global solution methods, namely SGs and HDMR. These methods have become an important and widely used tool in economics and finance for solving high-dimensional models, and are broadly applicable in both discrete and continuous time frameworks, as summarized in the recent review by \cite{Brumm2022}.\footnote{For an extensive review of numerical techniques for dynamic models, see the handbook chapters by \citealp{MALIAR2014325} and~\citealp{cai2014advances}.} In what follows, we present a brief, though by no means exhaustive, overview of the diverse applications within these fields.

The introduction of global sparse grids (GSGs; cf. Figure~\ref{fig:LitRev}), specifically the Smolyak sparse grid, into economics was pioneered by \citet{kruegerkubler_2004,kruegerkubler_2006}, focusing on discrete-time overlapping generations (OLG) models. \citet{fernandez-villaverde_2015} further utilized these GSGs to explore non-linear dynamics in a New Keynesian model constrained by a zero lower bound on nominal interest rates. Enhancements to the Smolyak method for better economic model performance were made by \citet{juddetal_2014}.
Local sparse grids (LSGs; cf. Figure~\ref{fig:LitRev}) entered economic analysis through the work of \citet{brummscheidegger_2017}, who applied these grids to IRBC models with irreversible investment and menu-cost models. Moreover, they introduced adaptivity, which adds a second layer of sparsity, as grid points are added only where they are most needed. \citet{brummetal_2017} and \citet{brumm2024public} employed LSGs for solving calibrated OLG models with aggregate shocks, while \citet{usui_2019_wp} used them to analyze rare natural disasters and adaptation in a dynamic stochastic economy.
Recent developments include the introduction of continuous-time adaptive LSG methods in macroeconomics by \citet{GarckeRuttscheidt2019}, particularly for heterogeneous agent models. \citet{Schaab_2020} utilized LSGs to solve HJB type equations, examining the interplay between micro and macro uncertainties in a heterogeneous agent New Keynesian model, accounting for aggregate risk, counter-cyclical unemployment, and monetary policy constraints.
In the realm of finance, LSGs have been applied both in discrete and continuous time to tackle high-dimensional option-pricing problems, as demonstrated by \citet{Reisinger:2007:EHA:1272826.1272843,Bungartz20123741,scheidegger_traccani_2018}, and to address dynamic portfolio choice models with transaction costs as per \citet{schober2021solving}.
Further financial applications include the integration of likelihood functions on GSGs by \citet{heiss2008likelihood}, the embedding of GSGs in Bayesian estimation frameworks by \citet{winschel_kraezig_2010}, and the application of LSGs in the context of the generalized method of moments by \citet{GilchGriebelOettershagen2021}.
~\cite{Young01022011} have used ideas from the HDRM literature in the context of splines to estimate linear models, whereas~\cite{doi:10.1137/21M1392231} combined SGs with HDMR to solve dynamic stochastic models containing up to 300 continuous states.
Finally, numerous high-performance, user-friendly open-source SG implementations are available in various programming languages. Among the most popular are \textit{SG++}\footnote{see~\url{https://sgpp.sparsegrids.org}}, the sparse grids Matlab kit\footnote{see~\url{https://sites.google.com/view/sparse-grids-kit}}, spinterp\footnote{see~\url{https://people.math.sc.edu/Burkardt/m_src/spinterp/spinterp.html}}, and TASMANIAN\footnote{see~\url{https://tasmanian.ornl.gov}}. In our numerical experiments within Dynare, we utilize TASMANIAN, as it is actively developed and maintained by Oak Ridge National Laboratory.

\section{Solving Dynamic Models with Sparse Grids and HDMR}
\label{Sec:dynamicModels}

To demonstrate the capabilities of embedding SGs and HDMR within Dynare to compute global solutions to a very broad range of discrete-time dynamic stochastic models, we begin by formally characterizing the class of models we aim to solve. 
Section~\ref{sec:2.1} presents the general structure common to many infinite-horizon, discrete-time dynamic stochastic economic models. 
Next, Section~\ref{sec:2.2} introduces a specific benchmark example, namely the IRBC model (see, e.g., \citealp{DenHaan2011175}).
Finally, in Section~\ref{sec:2.3}, we demonstrate how to iteratively compute global solutions for such dynamic models via time iteration, using the IRBC model as a guiding example.

\subsection{Abstract Model Formulation} 
\label{sec:2.1}

Let $\bold{x}_{t}\in X \subset \mathbb{R}^d$ denote the state of the economy at time $t\in \mathbb{N}$. 
We define the actions of all agents in the economy through a policy function $p: X \to Y$, 
where $Y$ represents the space of all possible policies. 
The evolution of the state $\bold{x}_t$ from period $t$ to $t+1$ follows the state transition
\begin{align}\label{eq:1}
	\bold{x}_{t+1} \sim \mathcal{D}(\cdot \mid \bold{x}_t, p(\bold{x}_t)),
\end{align}
where the distribution $\mathcal{D}(\cdot)$ is model-specific. 
The optimal policy function $p(\cdot)$ is not known a priori and must satisfy the 
period-to-period equilibrium conditions $E(\cdot)$. Specifically, the policy is time-invariant, so
\begin{align}\label{eq:2}
	\mathbb{E} \Bigl[ E\bigl(\bold{x}_t, \bold{x}_{t+1}, p(\bold{x}_t), p(\bold{x}_{t+1})\bigr) 
	\,\big\vert\, \bold{x}_t, p(\bold{x}_t) \Bigr] \;=\; 0 \quad \forall t,
\end{align}
where $\mathbb{E}[\cdot]$ is the expectation taken over the distribution in expression~\eqref{eq:1}.

This time-invariant policy $p(\cdot)$ can be determined via time iteration, 
by iterating directly on condition~\eqref{eq:2}. 
The time iteration algorithm~\citep{coleman1990} computes a recursive equilibrium of a dynamic economic model 
by starting with an initial guess for the policy function and iteratively refining it 
based on the model’s first-order conditions (e.g., \citealp{judd_1998}, Section~17.8). 
\paragraph{The Need for Global Solution Methods}
In many economic models, the equilibrium conditions $E$ exhibit substantial non-linearity due to concave utility and production functions or large shocks; moreover, (financial) frictions often imply non-differentiability. 
As a result, the optimal policy satisfying Equation~\eqref{eq:2} is generally non-linear and not necessarily smooth, necessitating function-approximation techniques suited to (high-dimensional) non-linear environments. Global solution methods, such as SGs and HDRM, are thus preferred over perturbation approaches, like those in the standard Dynare implementation~\citep{Adjemianetal2024}.

\subsection{International Real Business Cycle Model} 
\label{sec:2.2}

The IRBC model is a widely used benchmark for studying methods that solve high-dimensional dynamic stochastic models, 
because its dimensionality scales linearly with the number of countries. 
In the following, we present the equations of the IRBC model, which formally correspond to the abstract formulation given in Equation~\eqref{eq:2}.
A detailed discussion of the model itself is beyond the scope of this paper; 
the interested reader is referred to 
\citealp{DenHaan2011175} and  \citealp{brummscheidegger_2017} for derivations. 
Here, we focus on the explicit set of non-linear equations to be solved at many points in the state space 
in order to construct the optimal policy function $p(\cdot)$.

The IRBC model is dynamic and stochastic with a $(d=2N)$-dimensional state space, 
where $N \in \mathbb{N}_{+}$ is the number of countries. 
The state variables are
\begin{equation}\label{eq:3}
    \bold{x}_{t} = \bigl(a_{t}^1,\ldots,a_{t}^N,\;k_{t}^1,\ldots,k_{t}^N \bigr) \in \mathbb{R}^{2N},
\end{equation}
where $a_{t}^n$ and $k_{t}^n$ represent the productivity and capital stock of country $n\leqslant N$, respectively. 
The policy function $p:\mathbb{R}^{2N}\rightarrow\mathbb{R}^{N+1}$ 
maps the current state $\bold{x}_t$ into the next period,
\begin{equation}\label{eq:3.1}
     p(\bold{x}_t)=\bigl(k_{t+1}^1,\ldots,k_{t+1}^N,\;\lambda_t \bigr),
\end{equation}
where $\lambda_t$ is the multiplier for the aggregate resource constraint. 
For optimality, the IRBC policy must satisfy the following $N+1$ non-linear equations:
\begin{align}\label{eq:4}
	&\lambda_t \frac{\partial q^n_t\bigl(k^n_t,k^n_{t+1}\bigr)}{\partial k^n_{t+1}}
	\;-\;
	\beta \mathbb{E}_t \Biggl[
	\lambda_{t+1} 
	\frac{\partial\bigl(y^n(a^n_{t+1},k^n_{t+1}) - q^n(k^n_{t+1},k^n_{t+2})\bigr)}{\partial k^n_{t+1}}
	\Biggr] = 0 
	\quad \forall n,  
	\\
 	&\sum_{n=1}^N y^n\bigl(a^n_t,k^n_t\bigr) 
	\;-\; 
	\biggl(\frac{\lambda_t}{\tau^n}\biggr)^{-\gamma^n} 
	\;-\;
	q^n\bigl(k^n_t,k^n_{t+1}\bigr) \;=\; 0. 
	\nonumber
\end{align}
The parameters $\tau^n$ and $\gamma^n$ are model-specific, while 
$y^n(\cdot)$ and $q^n(\cdot)$ denote the production and convex adjustment-cost functions, respectively.
The complete parameterization of the model is given in \cite{brummscheidegger_2017}, Table~2.

\subsection{Time Iteration} 
\label{sec:2.3}

We next describe how the time iteration algorithm~\citep{coleman1990} is implemented in Dynare to iteratively solve dynamic models expressed abstractly as expression~\eqref{eq:2}, 
such as the IRBC model introduced in Section~\ref{sec:2.2}. 
Algorithm~\ref{alg:TimeIteration} illustrates how to solve the non-linear system \eqref{eq:4} 
using an interpolation/approximation, sloppily denoted as $\mathcal{I}$, of the previous policy guess, denoted by $\mathcal{I}p_0(\cdot)$. 
Specifically, we approximate the terms 
$\bigl(k_{t+2}^1,\ldots,k_{t+2}^N, \lambda_{t+1}\bigr)$ 
by $\mathcal{I}p_0\bigl(\bold{x}_{t+1}\bigr)$ and subsequently solve for the $N+1$ unknowns using a non-linear solver, such as \texttt{IPOPT}~\citep{Wachter:2006:IIF:1107694.1107695}.
Starting with an initial policy guess $\mathcal{I}p_{\text{guess}}$, which may be obtained from a linearized Dynare solution, the procedure iteratively refines the solution until it satisfies a specified tolerance, \texttt{tol}, measured, for instance, as mean squared error (cf. Sections~\ref{sec:SG_results} and~\ref{sec:HDMR_results} below).
In each iteration, a high-dimensional, non-linear policy is updated using an appropriate 
approximation scheme, such as SGs or HDMR (as discussed in this paper). 
For further implementation details, see \cite{brummscheidegger_2017}.
%
\begin{algorithm}[ht]
\caption{Time Iteration Algorithm.}
\label{alg:time_iteration}
\begin{algorithmic}[1]  
\Require $\mathcal{I}p_{\text{guess}}, \texttt{tol}$
\State $\mathcal{I}p \leftarrow \mathcal{I}p_{\text{guess}}$
\Repeat
  \State $\mathcal{I}p_0 \leftarrow \mathcal{I}p$
  \State construct $\mathcal{I}p$ by solving Eq.~(5) given $\mathcal{I}p_0$
\Until $\|\mathcal{I}p - \mathcal{I}p_0\| < \texttt{tol}$
\end{algorithmic}
\label{alg:TimeIteration}
\end{algorithm}

\section{Numerical Function Approximation}
\label{sec:approximator}

Dynamic stochastic models, solved via time iteration (see Sections~\ref{sec:2.2} and~\ref{sec:2.3}), necessitate repeated approximations and interpolations of high-dimensional, non-linear functions. In this section, we demonstrate that these tasks can be efficiently addressed using SGs or by combining SGs with HDMR, an approach we term, as mentioned before, \emph{dimension-decomposed sparse grids} (DDSG). Section~\ref{sec:3.1} provides a concise overview of SG techniques (e.g.,~\citealp{Bungartz.Griebel:2004,pflueger10spatially}), Section~\ref{sec:3.2} reviews HDMR and DDSG, and Section~\ref{sec:analytical_examples} presents analytical examples to build intuition.
Throughout, we adopt the notation established in~\cite{eftekharietal_2017}.


\subsection{Adaptive Sparse Grids} 
\label{sec:3.1}


%
\begin{figure}[t]
    \centering
    \includegraphics[width=0.99\textwidth]{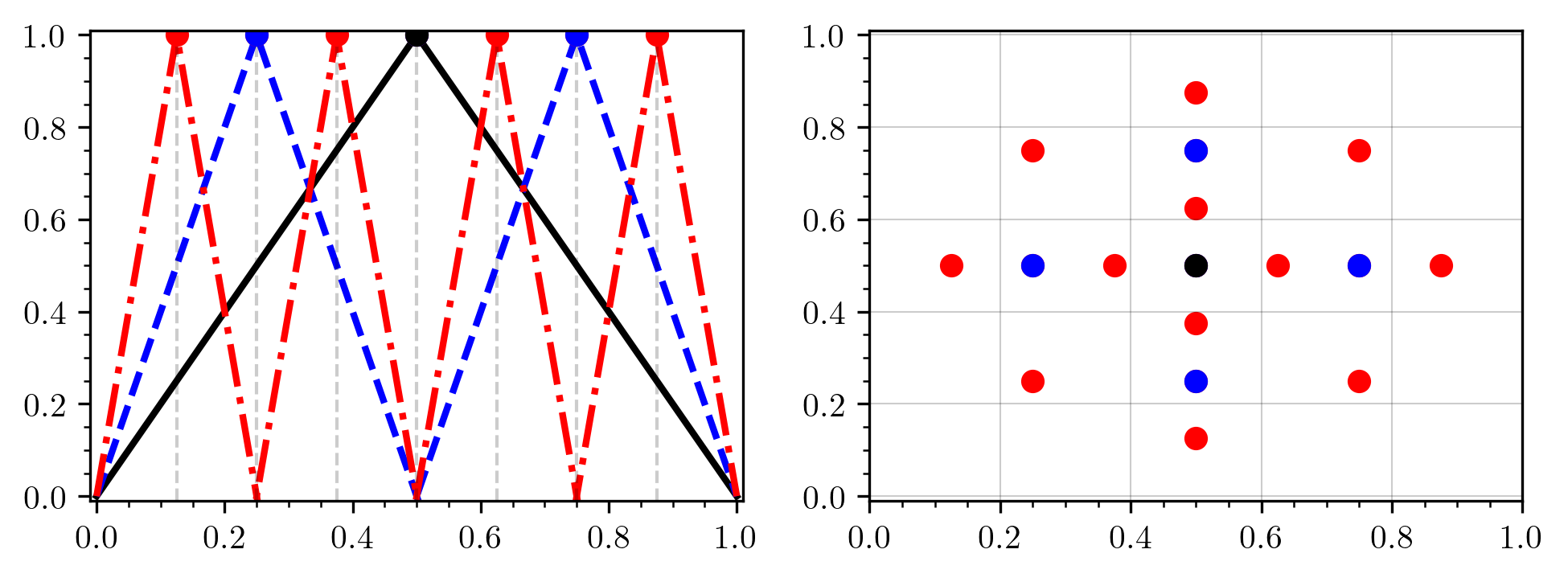}
    \caption{
Left Panel: Hierarchical basis functions at refinement levels $\ell=1$ (solid black), $\ell=2$ (dashed blue), and $\ell=3$ (dash-dotted red). 
Right Panel: A two-dimensional SG with corresponding refinement levels and colors.
}
\label{fig:sg}
\end{figure}
%
We aim to approximate the policy function, where each policy is represented as $f: \Omega \to \mathbb{R} $, with $ \Omega = \mathbb{R}^d $ and $ \bb{x} \subset \Omega $ rescaled to the unit domain $ [0,1]^d$.
In our case $d$ the number of continuous state variables in the economic model of interest, a potentially large number.\footnote{For illustration and notational brevity, we assume zero boundary conditions. For a discretization scheme with nonzero boundary conditions—such as the Clenshaw-Curtis sparse grid setting employed in all our numerical experiments, see~\cite{stoyanov2015tasmanian} and \cite{pflueger10spatially}.}
In one dimension, the unit domain $[0,1]$ can be discretized with grid spacing $h_l = 2^{-l}$, grid points at $x_{l,i} = i \cdot h_l$, where $i \in \{1, \dots, 2^l\}$ and refinement level $l \in \mathbb{N}_+$.
The univariate basis functions for this discretization are defined as
\begin{align} \label{eq:3.1}
\phi_{l,i}(x) = \max \left(1 - \frac{1}{h_l}\,\bigl|x - x_{l,i}\bigr|,\,0\right),
\end{align}
which have support $[x_{l,i} - h_l,\, x_{l,i} + h_l]$.
In the left panel of Figure~\ref{fig:sg}, we illustrate a one-dimensional hierarchical piecewise linear basis function.
We extended the basis function to a $d$-dimensional domain by first introducing the multi-indices for the refinement level $\bb{l}=({l}_1,\ldots, {l}_d) \in \mathbb{N}_{+}^d $ and grid index $\bb{i}=({i}_1,\ldots, {i}_d)$ with the corresponding grid point $\bb{x}_{\bb{l},\bb{i}} = (x_{l_1,i_1}, \dots, x_{l_d,i_d})$.
The $d$-dimensional basis functions are defined as
\begin{align} \label{eq:3.2}
\phi_{\bb{l},\bb{i}}(\bb{x}) = \prod_{j=1}^d \phi_{l_j,i_j}(x_j).
\end{align}
We define the hierarchical index sets $\bb{I}_{\bb{l}}$, and the corresponding hierarchical subspace $W_{\bb{l}}$ as
\begin{align} \label{eq:3.3}
\bb{I}_{\bb{l}} = \left\{\bb{i} : 0 < i_j < 2^{l_j},\; i_j\, \text{odd},\; 1 \leqslant j \leqslant d \right\}, 
\quad \text{and} \quad
W_{\bb{l}} = \operatorname{span}\left\{ \phi_{\bb{l},\bb{i}} : \bb{i} \in \bb{I}_{\bb{l}} \right\}.
\end{align}
The restriction to odd indices $i_j$ ensures that the supports of the basis functions are disjoint and collectively cover $[0,1]^d$.
For the space of piecewise linear functions,
\begin{align}\label{eq:3.4}
    V_\ell = \bigoplus_{ \norm{\mathbf{l}}{\infty} \leqslant \ell} W_{\mathbf{l}},
\end{align}
we can construct a corresponding equidistant Cartesian grid, also called a \emph{full grid}, with $M_{\ell}=2^{\ell}$ number of grid points in each dimension, where $\ell$ denotes the \emph{maximum refinement level}.

Although the $L_2$ interpolation error is of order $\mathcal{O}(M_\ell^{-2})$, the total number of grid points is $\mathcal{O}(M_\ell^d)$, effectively rendering this approach impractical for high-dimensional functions.
SG mitigates the curse-of-dimensionality by retaining only the most significant hierarchical subspaces. 
Formally, the SG space is defined as
\begin{align} \label{eq:3.5}
V^{\text{SG}}_\ell = \bigoplus_{\|\bb{l}\|_1 \le \ell + d - 1} W_{\bb{l}}.
\end{align}
The SG interpolation of $f$ at a point $\bb{x}$ with a maximum refinement level $\ell$ is defined as
\begin{align} \label{eq:3.6}
\mathcal{I}_\ell f(\bb{x}) := \sum_{\|\bb{l}\|_1 \le \ell + d - 1} \sum_{\bb{i} \in \bb{I}_{\bb{l}}} \alpha_{\bb{l},\bb{i}}\,\phi_{\bb{l},\bb{i}}(\bb{x}),
\end{align}
where $\alpha_{\bb{l},\bb{i}} \in \mathbb{R}$ (hierarchical surpluses).
In the right panel of Figure~\ref{fig:sg}, we illustrate the grid points for a two-dimensional SG.
For functions with bounded mixed second derivatives, the SG interpolation error is of the order $\mathcal{O}(M_\ell^{-2} (\log M_\ell)^{d-1})$, while the number of grid points is $\mathcal{O}(M_\ell (\log M_\ell)^{d-1})$---substantially less grid points than the full grid in high dimensions.
For further details on SGs, including the error analysis and the computation of hierarchical surpluses $\alpha_{\bb{l},\bb{i}}$, we refer the reader to~\cite{Bungartz.Griebel:2004} and references therein.
Finally, quadrature on SGs, denoted as
\begin{align}\label{eq:3.7}
     \mathcal{Q}_\ell f = \sum_{\|\bb{l}\|_{1} \le \ell+d-1} \sum_{\bb{i} \in \bb{I}_{\bb{l}}} \alpha_{\bb{l},\bb{i}} \int \phi_{\bb{l},\bb{i}}(\bb{x})\,d\bb{x},
\end{align}
can be efficiently evaluated by integrating the basis functions defined in Equation~\eqref{eq:3.2}. 

If the function to be approximated has sharp local features, steep gradients, or non-differentiabilities---for example, in economic models with occasionally binding constraints---the assumed condition of bounded second-order mixed derivatives is not satisfied. 
As a result, significantly higher SG refinement levels (and thus more grid points) may be required, limiting the applicability of SG in many high-dimensional problems of interest.
In such scenarios, an \emph{adaptive} refinement procedure can be employed to preserve efficiency~\citep[e.g.,][]{pflueger10spatially,brummscheidegger_2017}. 
In this adaptive approach, one monitors the magnitude of each hierarchical surplus $\alpha_{\bb{i},\bb{l}}$, which indicates the local irregularity of the function at $\bb{x}_{\bb{i},\bb{l}}$. 
Let $\epsilon_\gamma \in \mathbb{R}_+$ be a predefined refinement threshold, and define a refinement criterion $g$.
If $g(\alpha_{\bb{i},\bb{l}}) \leqslant \epsilon_\gamma$, then the point $\bb{x}_{\bb{i},\bb{l}}$ is considered \emph{insignificant} and no further refinement is performed at that location. 
Due to the hierarchical structure of the grid, excluding $\bb{x}_{\bb{i},\bb{l}}$ automatically excludes any descendant points at higher refinement levels. 
In many practical applications, the commonly used refinement criterion is
\begin{align}\label{eq:10}
g\bigl(\alpha_{\bb{i},\bb{l}}\bigr):=\bigl|\alpha_{\bb{i},\bb{l}}\bigr|,
\end{align}
although alternative criteria may also be applied \citep[e.g.,][]{stoyanov2015tasmanian}.

SGs are, as mentioned above, highly effective for approximating moderately high-dimensional functions (say, \(d<20\)); however, they can become computationally prohibitive as dimensionality or complexity increases (cf.~\cite{brummscheidegger_2017}). To address this limitation, we now turn our attention to HDMR and DDSG.

\subsection{Dimensional Decomposition}
\label{sec:3.2}

In this section, we follow the approach of~\cite{hooker_2007} and~\cite{Rabitz1999GeneralRepresentations} to illustrate the fundamental steps for constructing a dimensional decomposition (DD).
As in the previous sections, we consider a scalar-valued function $f(\bb{x})$, where $\bb{x} \in [0,1]^d$.\footnote{
The presented method is not limited to scalar-valued functions but can be directly extended to vector-valued functions. The scalar formulation is used solely for notational simplicity.
}
Denote $\bb{u} \subseteq \mathcal{S}=\{1,2,\ldots,d\}$ as a \emph{component index}, and $f_{\bb{u}}(\bb{x_u})$ as a \emph{component function}, where $\bb{x}_{\bb{u}}$ is a vector that consists of the values $x_i$ for $i \in \bb{u}$.
The function $f(\bb{x})$ can be expressed as the expansion 
\begin{align} \label{eq:13} 
  f(\bb{x}) &= \sum_{\bb{u} \subseteq \mathcal{S}} f_{\bb{u}}(\bb{x_u}).
\end{align}
This representation of the function is referred to as HDMR.
The terms in the summation are classified by the \emph{expansion order} $k:=\card{\bb{u}}$, which corresponds to the dimension of the input vector $\bb{x}_{\bb{u}}$. 
Expressed explicitly, the function can be decomposed as
\begin{align} \label{eq:13a} 
  f(\mathbf{x}) &= f_{\emptyset}+\sum_{1\leqslant i \leqslant d} f_{i}({x}_i) +\sum_{1\leqslant i < j \leqslant d } f_{i,j}({x}_i,{x}_j) + \ldots + f_{1,2,\ldots,d}({x}_1,{x}_2,\ldots, {x}_d ). 
\end{align}
Here, $f_{\emptyset}$ is a constant term corresponding to the zeroth-order contribution; the functions $f_i$, with $1\leq i\leq d$, are univariate (first-order) contributions; the functions $f_{i,j}$ represent bivariate (second-order) contributions; and so forth, culminating in the $d$th-order contribution $f_{1,2,\dots,d}$.
In its complete form, this decomposition is exact, since the final term captures all interactions among the input variables.
The principal advantage of the DD approach becomes evident when the high-dimensional function \( f \) can be well approximated by truncating the expansion in expression~\eqref{eq:13} at a maximum order \(\ekk \ll d\).

Among the various formulations of HDMR, \emph{cut-HDMR} and \emph{ANOVA-HDMR} are particularly prominent (e.g., \citealp{RABITZ199911} and \citealp{genyuan}).\footnote{For a comprehensive discussion of alternative cut-HDMR variants, including RS-HDMR, mp-cut-HDMR, Multicut-HDMR, and lp-RS-HDMR, see~\cite{Li2012GeneralVariables}.}
We focus specifically on cut-HDMR, as it more closely aligns with the goals of our application. In contrast to ANOVA-HDMR, which requires high-dimensional numerical integration, cut-HDMR relies solely on direct function evaluations (see also \cite{doi:10.1137/21M1392231}).

Let $w(\bb{x}) = \prod_{i=1}^d w_i({x}_i)$ be a product measure, with $w_{i}({x}_i)$ having a unit volume.
By sequentially ascending through the expansion orders, starting from the zeroth order, the optimally and uniquely defined HDMR component function
\begin{align} \label{eq:ddsg2.2.4} 
   f_{\bb{u}}(\bb{x}_{\bb{u}}) =  &\argmin{g_{\bb{u}}} \int \left (  \sum_{\bb{u} \subseteq \mathcal{S}} g_{\mathbf{u}}(\mathbf{x_u}) - f(\mathbf{x}) \right )^2 w(\bb{x}) d \bb{x}, \\
     &\text{subject to} \;  \int g_\mathbf{u} (\mathbf{x_{u}}) w_{i}(\bb{x}_i) d {x}_i=0, \;\; \forall i \in \mathbf{u},\nonumber 
\end{align}
will only be dependent on lower-order component functions $f_{\bb{v}}(\bb{x}_{\bb{x}})$ for $\bb{v} \subset \bb{u}$.
This is particularly important because the contrary would eliminate any reduction in dimensionality.
This attribute is a result of the orthogonality condition imposed in Equation~\eqref{eq:ddsg2.2.4}~\citep{Rabitz1999GeneralRepresentations,hooker_2007}.
The cut-HDMR component functions are defined using the Dirac measure,
\begin{align}\label{eq:ddsg2.2.5}
 w(\bb{x})\,d\bb{x} = \prod_{i=1}^d \delta({x}_i - \bar{x}_i)\,d{x}_i,  
\end{align}
where $\delta(\cdot)$ denotes the Dirac delta function, and $\bar{\bb{x}}=(\bar{x}_1,\dots,\bar{x}_d)$ is a reference point known as the \textit{anchor point}.  
As shown by~\cite{Sobol2003TheoremsRepresentation}, a suitable anchor point should satisfy
\begin{align}\label{eq:ddsg2.2.8}  
\bar{\bb{x}} \approx \argmin{\bb{z}} \|f(\bb{z}) - \mathbb{E}[f(\mathbf{x})]\|_1,
\end{align}
and can be selected by sampling $\bbb{x}$ so that $f(\bbb{x})$ is close to the mean of the function.
Evaluating expression~\eqref{eq:ddsg2.2.4}, the cut-HDMR component functions are defined as
\begin{align}\label{eq:ddsg2.2.6} 
    f_{\bb{u}}(\bb{x_u}) = \sum_{\bb{v} \subseteq \bb{u}}(-1)^{\card{\bb{u}}-\card{\bb{v}}}f(\bb{x})\eval{\bb{x}}{\bar{\bb{x}} \backslash \bb{x_v}}, \quad \text{with}\, f_\emptyset &= f(\bar{\bb{x}}) .
\end{align}
We use the notation $\bb{x} = \bar{\bb{x}} \backslash \bb{x_v}$ to refer to assigning $\bb{x}$ the values of $\bar{\bb{x}}$ but excluding the indices of $\bb{v}$.
For example, given $\bb{x}=({x}_1,{x}_2,{x}_3)$, then $\bbb{x} \backslash \bb{x}_{1,2}=({x}_1,{x}_2,\bar{x}_3)$.

\begin{figure}[t!] 
	\centering
\includegraphics[width=0.99\textwidth]{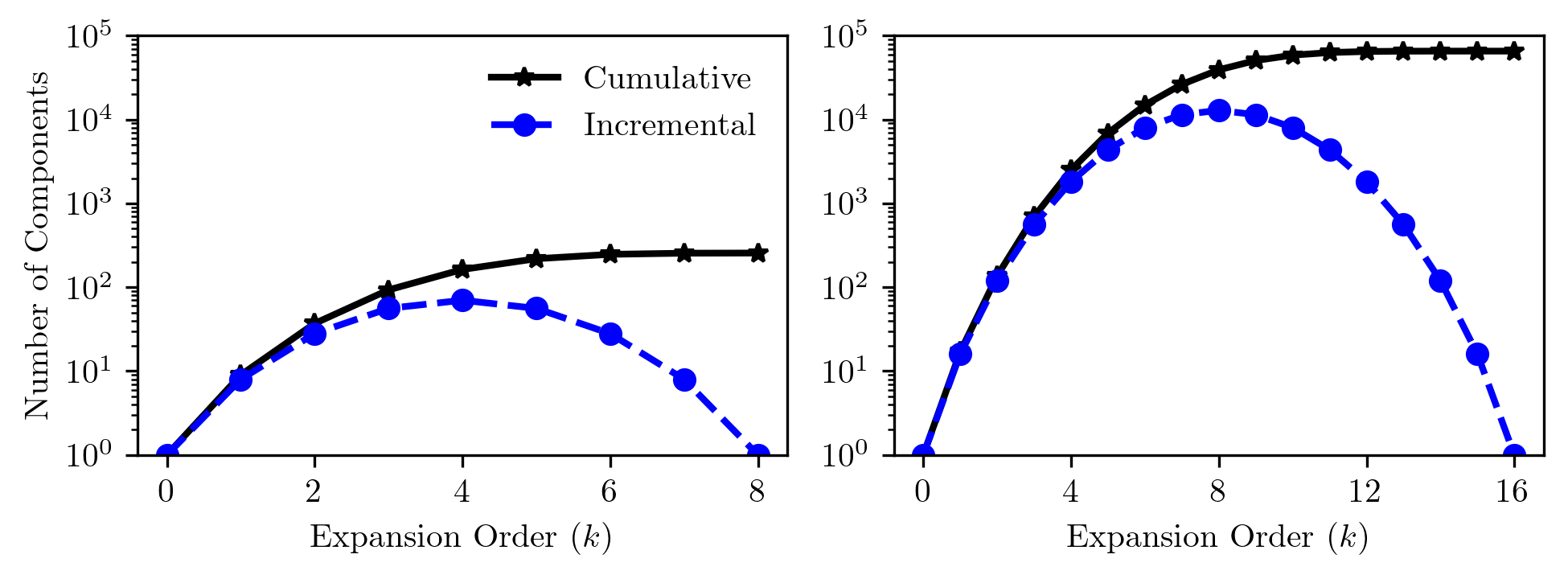}
\caption{The number of DD component functions, both incremental and cumulative, as a function of expansion order for an \(8\)-dimensional (left panel) and a \(16\)-dimensional function (right panel).
The number of component functions increases significantly with both expansion order and dimensionality.
}
\label{fig:expansion}
\end{figure}\noindent
%
Performing the full expansion in expression~\eqref{eq:13} up to order \( k = d \) is computationally infeasible for problems of nontrivial dimensionality, as it simply reconstructs the original \( d \)-dimensional function, thereby negating the intended efficiency gains from dimensional reduction.
Moreover, the number of component functions increases combinatorially with the expansion order \( k \):
\begin{align}\label{eq:19}
\sum_{j=0}^{k} \frac{d!}{(d-j)!j!},
\end{align}
which poses significant computational challenges even for moderately sized problems (e.g., 8- or 16-dimensional), as illustrated in Figure~\ref{fig:expansion}.
Therefore, truncating the expansion to a maximum order \(\ekk \ll d\) is imperative; moreover, the optimal selection of \(\ekk\) is inherently problem-dependent.
\begin{figure}[t!]
\centering
\begin{forest}
for tree={
    grow=south
}
[$f_{1234}$,rectangle, draw, fill=gray!20,
    [$f_{123}$,rectangle, draw, fill=gray!20,
        [$f_{12}$,rectangle, draw, fill=gray!20,
            [$f_{1}$,rectangle, draw,
                        [$f_{\emptyset}$,rectangle, draw]
            ]
            [$f_{2}$,rectangle, draw]
        ]
        [$f_{13}$,rectangle, draw,
            [$f_{3}$,rectangle, draw]
        ]
        [$f_{23}$,rectangle, draw]
    ]
    [$f_{124}$,rectangle, draw, fill=gray!20,
        [$f_{14}$,rectangle, draw,
            [$f_{4}$,rectangle, draw]        
        ]
        [$f_{24}$,rectangle, draw]
    ]
    [$f_{134}$,rectangle, draw,
        [$f_{34}$,rectangle, draw]
    ]
    [$f_{234}$,rectangle, draw]
]         
\end{forest}
\caption{
Visualization of the component functions summation in Equation~\eqref{eq:13} of a four-dimensional function with active dimension criteria, where the component index $\bold{u}=\{1,2\}$ is deemed insignificant.
For clarity, repeated cells are omitted. 
All component functions that are supersets of $\bold{u}=\{1,2\}$ are excluded from the summation (shown in gray).
}
\label{fig:prune}
\end{figure}
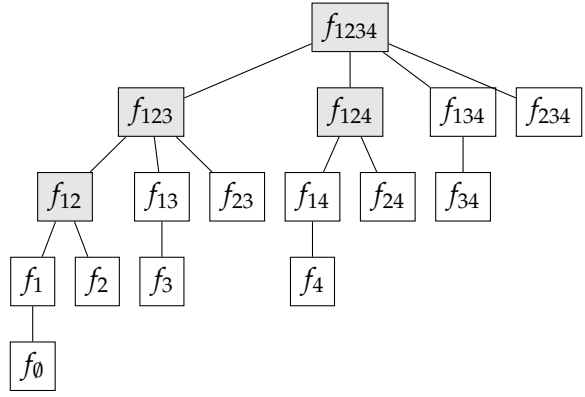
%

\paragraph{Truncating the HDMR Expansion}
Two adaptive criteria have proven particularly effective for truncating the expansion in Equation~\eqref{eq:13}, especially in economic applications~\citep{doi:10.1137/21M1392231}. These criteria involve: (i) assessing the relative importance of individual component functions, and (ii) evaluating the incremental benefit of proceeding to higher expansion orders.

The first criterion, termed \textit{active dimension selection}, evaluates the significance of each component function by comparing the norm of its integral to the norm of the cumulative integral of all previously computed lower-order component functions. Component functions whose integrals fall below a specified threshold are discarded, together with all supersets of their corresponding indices (see Figure~\ref{fig:prune} for a visual representation).
The second criterion, known as the \textit{expansion criterion}, examines convergence across expansion orders by quantifying the incremental changes in the integral values of the component functions. Expansion is terminated once these incremental changes fall below a predefined convergence threshold.
Both criteria exploit the hierarchical structure of the expansion, thereby enhancing computational efficiency by reusing previously computed integrals.\footnote{Starting from lower-order terms, numerical integration is performed in lower-dimensional spaces, thus mitigating the computational challenges associated with high-dimensional integration.}

\paragraph{Dimensional Decomposition With Adaptive Sparse Grids}
\label{sec:3.3}
%
The abstract formulation of DD provided above has been described independently of the numerical methods for interpolation and quadrature. 
However, practical implementation requires efficient numerical schemes, particularly at higher DD expansion orders, due to the curse of dimensionality.
SGs are particularly well suited for numerical methods for DD because of two key attributes: (i) SGs are effective in approximating high-dimensional functions with localized features, and (ii) they capability for efficient numerical integration (see Equation~\eqref{eq:13}). 
This combined approach is referred to as DDSG.
Following the previously established exemplar of the \( d \)-dimensional function \( f \), the DDSG function is expressed as
\begin{align}\label{eq:23} 
	f(\mathbf{x}) &\approx \sum_{ \substack{ \bb{u} \subseteq \mathcal{S} \\  \card{\bb{u}} \leqslant \ekk  } } \sum_{\bb{v} \subseteq \mathbf{u}}(-1)^{\card{\bb{u}}-\card{\bb{v}}} \mathcal{I}_{\ell}f(\bb{x})\eval{\bb{x}}{\bar{\bb{x}} \backslash \mathbf{x_v}} ,   \\
	&\approx \sum_{ \substack{ \bb{u} \subseteq \mathcal{S} \\  \card{\bb{u}} \leqslant \ekk  } } \sum_{\bb{v} \subseteq \bb{u}} (-1)^{\card{\bb{u}}-\card{\bb{v}}} \hspace{-1em} \sum_{\norm{\bb{k}}{1} \leq \ell+d-1} \sum_{\bb{i} \in \bb{I_k}} \alpha_{\bb{i},\bb{k}}\, \phi_{\bb{i},\bb{k}}(\bb{x})\eval{\bb{x}}{\bar{\bb{x}} \backslash \bb{x_v}} \nonumber 
\end{align}
which represents a nested summation over a series of \(|\bb{v}|\)-dimensional (adaptive) SGs.
The DDSG-based quadrature procedure adopts a similar structure~\citep{doi:10.1137/21M1392231}.
Note that we impose a strict maximum expansion order, which can be effectively combined with the adaptive criteria discussed above (e.g., active dimension selection and convergence criteria).


\subsection{Analytical Examples}
\label{sec:analytical_examples}
In this section, we present an analytical example to highlight the potential efficiencies and drawbacks of DDSG compared to SG. Specifically, we aim to demonstrate that, when a (latent) additively separable structure exists, DDSG can enhance the efficiency of SG by orders of magnitude, whereas in the absence of such a structure, SG outperforms DDSG.
%
\begin{figure}[t!] 
	\centering
\includegraphics[width=0.99\textwidth]{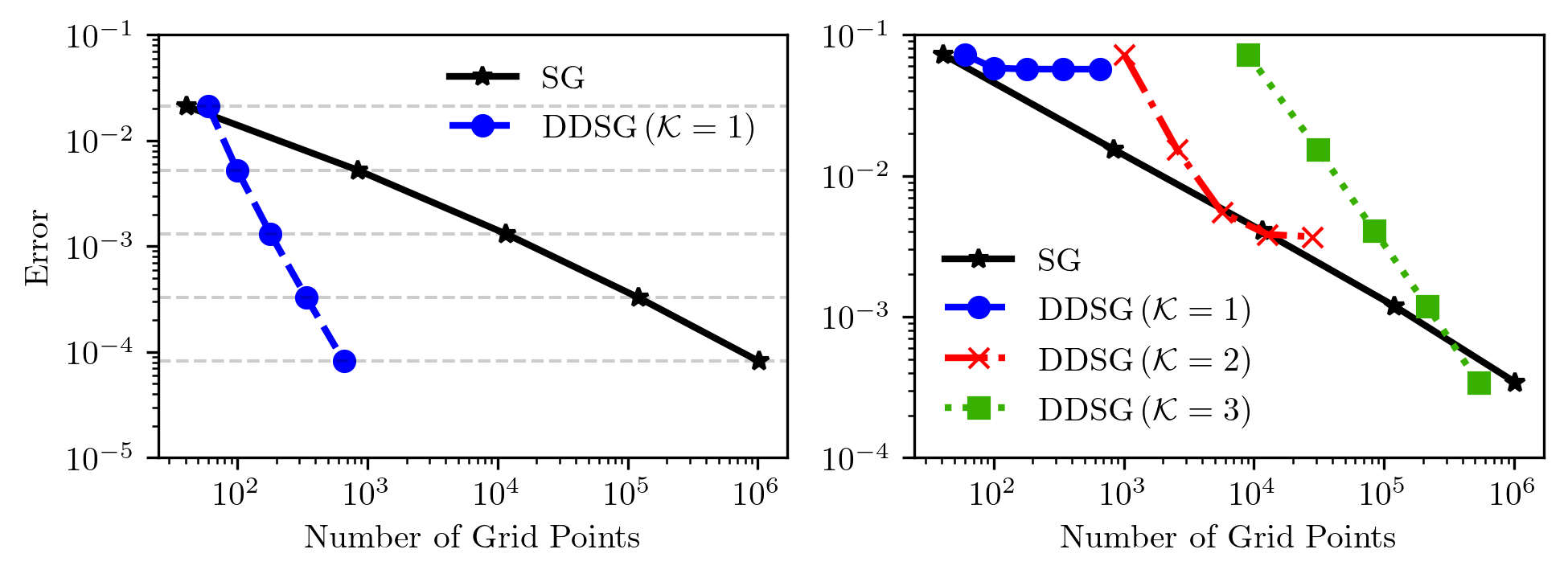}
\caption{Interpolation error for SG and DDSG at different maximum expansion orders \(\ekk\) for the test function given in Equation~\eqref{eq:test}, with coefficients (left panel) \(c = 1\) and (right panel) \(c = 3\). The approximation error is measured as the relative average error over \(1,000\) samples of \(\bb{x} \in [0,1]^d\), where $d=20$.}
\label{fig:test}
\end{figure}\noindent
%

We consider the following non-linear, analytical test function: 
\begin{align}
  f(\mathbf{x}) = \left( \sum_{i=1}^{d} \sin(x_i) \right)^c,
\label{eq:test}
\end{align}
where \(c = \{1, 2, \ldots\}\). 
This function is additively separable, and DD provides an exact decomposition with expansion order \(r = c\).
For any case where \(\ekk < c\), the DDSG approach introduces an approximation error that cannot be mitigated by increasing the refinement levels of the SG (applied to the lower-dimensional component functions).
In Figure~\ref{fig:test}, we compare the approximations generated by SG and DDSG for a \(20\)-dimensional example, evaluating the average relative interpolation error using $1,000$ random samples over the domain \(\bb{x} \in [0,1]^d\). 
Each marker in the plots represents a refinement level, ranging from \(\ell = 1, \ldots, 5\).  
In the left panel of the figure the test function uses \(c = 1\), meaning that DDSG with expansion order \(\ekk = 1\) achieves the same interpolation error as SG but with substantially fewer grid points.  
In the right panel of the figure the test function has \(c = 3\), where DDSG with expansion order \(\ekk = 1\) performs poorly compared to SG, and the minimum error remains relatively high regardless of the refinement level. 
Increasing the expansion order reduces the error but significantly increases the number of grid points due to the growing number of component functions.  
Only when DDSG has an expansion order of \(\ekk = c = 3\) does the decomposition become exact, allowing it to achieve the same error as SG.

Notice that for this test function, a simple log-transform can effectively render the function additively separable with a DD expansion order \(\ekk = 1\), regardless of the value of \(c\), a technique that could also be applied in the context of economic models.
While this approach may not be universally applicable, an appropriate transformation can significantly reduce DDSG approximation errors for non-additively separable functions.
Moreover, in real applications where the function is unknown, the effective expansion criterion outlined in Section~\ref{sec:3.2} can be used to implicitly deduce the separability of the unknown function (see~\citet{doi:10.1137/21M1392231} for more details).

\section{Dynare Syntax} 
\label{sec:syntax}

This section details the required modifications to Dynare's \texttt{.mod} files to accommodate the proposed generic global solution methods.\footnote{Notice that our code developments were implemented in Julia (i.e., \texttt{Dynare.jl}). However, the model parsing is independent of the programming language, so the tools developed can also be made available in Dynare environments such as \texttt{Matlab}.}
Code Listing~\ref{tab:code_listing} presents the \texttt{.mod} file command for Dynare that characterize the IRBC model (cf. Section~\ref{sec:2.2}), beginning with the specification of variables shared across countries. The subsequent declaration of variables, shock innovations, and parameters follows standard Dynare syntax.

We then leverage Dynare's macro-language to programmatically specify country-specific model variables, shocks, equations, and initial conditions. The macro-language extends Dynare's basic syntax by introducing conditionals, loops, display, and error directives, among other commands.\footnote{More details on Dynare's macro-language and macro-processor are available in \href{https://www.dynare.org/manual/the-model-file.html\#macro-processing-language}{Dynare's manual}.} The macro-processor expands these commands in the \texttt{.mod} file, generating a fully specified version that Dynare can process using its standard workflow. Since this macro expansion occurs as a pre-processing step, it ensures that the final file conforms to standard Dynare syntax. This approach simplifies the creation of \texttt{.mod} files by automatically handling repetitive variable definitions and equations, as demonstrated in the multi-country model considered in this study.

In our implementation, the \texttt{@\#define} command sets the macro-variable \texttt{N}, representing the number of countries in the model. We then define country-specific variables and parameters using a macro loop statement (\texttt{@\#for}-\texttt{@\#endfor}) combined with expression substitution, where \texttt{@\{j\}} is replaced by the current value of the macro-variable \texttt{j} in the \texttt{.mod} file. Specifically, for illustrative purposes, we fix the number of countries at \texttt{N} \(= 2\). The country-specific variables then include the capital levels (denoted by \texttt{k\_1} and \texttt{k\_2}) and log-productivity levels (denoted by \texttt{a\_1} and \texttt{a\_2}).
Similarly, \texttt{e\_1} and \texttt{e\_2} represent the country-specific components of the productivity innovations. The country-specific parameters comprise the utility function curvatures (\texttt{gamma\_1} and \texttt{gamma\_2}) and the welfare weights (\texttt{t\_1} and \texttt{t\_2}).

The \texttt{model} block also leverages the macro-language to programmatically declare country-specific optimality conditions for investment in capital and productivity processes using macro loop statements. In contrast to standard Dynare models—where exogenous variables (\texttt{varexo}) represent pure stochastic innovations—our approach distinguishes between stochastic innovations and shock-driven state variables. This distinction is made explicit in the \texttt{[preamble]} block, which groups all equations governing the processes of exogenous variables. In our setup, the productivity processes (\texttt{a\_1} and \texttt{a\_2}) are classified as exogenous because their evolution depends solely on their past values and stochastic innovations (\texttt{e} and \texttt{e\_j}), unlike the conventional Dynare approach in which only the innovations are exogenous and the full autoregressive process is treated as endogenous. The dedicated \texttt{[preamble]} block isolates the exogenous dynamics, thereby enhancing computational efficiency, particularly in numerical methods such as sparse grids. Moreover, our implementation requires that shock innovations be set to 1, so users must manually scale the innovations in the \texttt{[preamble]} equations by their respective standard errors. The \texttt{shocks} block then assigns the standard errors of country-specific shock innovations using a macro loop. Finally, the \texttt{initval} block specifies initial values for simulation or serves as initial guesses for non-linear solvers, with a macro loop setting the values for the country-specific capital and log-productivity levels.

Thus far, declarations (variables, shocks, parameters, and model equations) largely adhere to standard Dynare syntax, aside from a few modifications in the definition of exogenous variables. The macro-language is employed primarily to automate and structure repetitive components, thereby ensuring scalability as the number of countries increases. We further exploit the ability of the \texttt{.mod} file to interpret native Julia code by incorporating the SG routines \texttt{sparsegridapproximation} and \texttt{DDSGapproximation}.

We leverage the capability of the \texttt{.mod} file to interpret native Julia code by incorporating the SG routines \texttt{sparsegridapproximation} and \texttt{DDSGapproximation}. The function \texttt{sparsegridapproximation} implements an adaptive sparse grid method that iteratively refines the approximation grid to solve the model. Its accuracy and computational cost are primarily determined by several key parameters. In particular, the parameter \texttt{gridDepth} sets the initial grid depth, thereby influencing the number of interpolation points. SG refinements are controlled by \texttt{maxRef}—which specifies the maximum number of refinement steps—and by \texttt{surplThreshold}, which defines the surplus error threshold that triggers refinement. The time iteration process employs a convergence criterion specified by \texttt{tol\_ti}. Additionally, the parameter \texttt{polUpdateWeight} regulates the update of the policy function during iterations by assigning a weight to the newly computed policy relative to the previous iteration. Lower values of \texttt{polUpdateWeight} stabilize the updates, albeit at the potential cost of slower convergence.

The \texttt{DDSGapproximation} function extends the standard SG approach by employing the DDSG method, which restricts interactions among state variables to alleviate the curse of dimensionality. In addition to the parameters used in \texttt{sparsegridapproximation}, \texttt{DDSGapproximation} introduces \texttt{k\_max}, which specifies the maximum order of interaction terms in the DDSG decomposition. This parameter is pivotal in balancing computational efficiency with approximation accuracy.
%

\begin{tcolorbox}[
    colback=white, 
    arc=10pt, 
    boxrule=0.5pt, 
    breakable, 
    before skip=5pt, 
    after skip=5pt, 
    enlarge top at break by=0pt, 
    enlarge bottom at break by=0pt, 
    left=2mm, 
    right=2mm, 
    top=2mm, 
    bottom=2mm 
]
\scriptsize 
\begin{verbatim}
var lambda;
varexo e;
parameters kappa beta delta phi rho A sigE;

kappa = 0.36;
beta  = 0.99;  
delta = 0.01;
phi   = 0.5;
rho   = 0.95;
sigE = 0.01;
A = (1 - beta*(1 - delta))/(kappa*beta);

a_eis = 0.25;
b_eis = 1;

@#define N=2
@#for j in 1:N
var k_@{j} a_@{j};
  varexo e_@{j};
  parameters gamma_@{j} t_@{j};
  gamma_@{j} = a_eis + (@{j} - 1)*(b_eis - a_eis)/(@{N}-1);
  t_@{j} = A^(1/gamma_@{j});
@#endfor

model;
  @#for j in 1:N
    lambda*(1 + phi*(k_@{j}/k_@{j}(-1) - 1))
      = beta*lambda(+1)*(exp(a_@{j}(+1))*kappa*A*k_@{j}^(kappa - 1)
        + 1 - delta + (phi/2)*(k_@{j}(+1)/k_@{j} - 1)*(k_@{j}(+1)/k_@{j} + 1));
      [preamble]
      a_@{j} = rho*a_@{j}(-1) + sigE*(e + e_@{j});
  @#endfor
    exp(a_1)*A*k_1(-1)^kappa
  @#for j in 2:N
    + exp(a_@{j})*A*k_@{j}(-1)^kappa
  @#endfor
  =
  (lambda/t_1)^(-gamma_1) + k_1 - (1 - delta)*k_1(-1) + (phi/2)*k_1(-1)*(k_1/k_1(-1) - 1)^2
  @#for j in 2:N
  + (lambda/t_@{j})^(-gamma_@{j}) + k_@{j} - (1 - delta)*k_@{j}(-1)
             + (phi/2)*k_@{j}(-1)*(k_@{j}/k_@{j}(-1) - 1)^2
  @#endfor
  ;
end;

initval;
@#for j in 1:N
    k_@{j} = 1;
    a_@{j} = 0;
  @#endfor
  lambda = 1;
end;

steady;

shocks;
  var e; stderr 1;
  @#for j in 1:N
    var e_@{j}; stderr 1;
  @#endfor
end;

@#for j in 1:N
  limits!("k_@{j}", min = 0.8, max = 1.2);
  limits!("a_@{j}", min = -0.8*sigE/(1 - rho), max = 0.8*sigE/(1 - rho));
@#endfor

(SG_grid, sgws) = sparsegridapproximation(gridDepth=3,maxRef=0);
(DDSG_grid, ddsgws) = DDSGapproximation(gridDepth=3,maxRef=0,k_max=1);
\end{verbatim}
\label{tab:code_listing}
\end{tcolorbox}

\section{Results} 
\label{sec:results}

We now evaluate the performance of SGs and DDSG within Dynare for computing global solutions to our benchmark IRBC model. Section~\ref{sec:SG_results} reports SG results, while Section~\ref{sec:HDMR_results} examines the DDSG performance.\footnote{All tests presented in this section used a standard laptop with an Intel Core i9-12900H (14 cores, 20 threads, 2.5 GHz) and 64 GB of memory.} These experiments illustrate that global solutions of high-dimensional stochastic models can be computed both accurately and swiftly in Dynare without
any need to resort to high-performance computing and with only minimal modifications, thereby broadening the scope of applications available to the Dynare community. This section demonstrates that i) SG and DDSG operate reliably within Dynare, and ii) leveraging Dynare’s perturbation solution as an initial guess for the time iteration algorithm significantly reduced the time to solution. For clarity of exhibition, we disable SG adaptivity throughout this section and employ a fixed SG with a specified refinement level. Discussions on adaptivity and model-specific hyperparameter tuning for SGs and DDSG are beyond the scope of this paper and are addressed in~\cite{brummscheidegger_2017} and~\cite{doi:10.1137/21M1392231}, respectively.

\subsection{Solving the IRBC model with SGs and Dynare}
\label{sec:SG_results}

To systematically evaluate how accuracy varies with grid refinement \(\ell\) and problem dimensionality, we examine SG solutions for the IRBC model, following the performance metrics discussed in \cite{juillard_et_al_2011} and \cite{brummscheidegger_2017}.
The time iteration algorithm (Algorithm~\ref{alg:time_iteration}), initialized with Dynare’s first-order perturbation solution, was executed until either an accuracy of \(\texttt{tol}=1\cdot 10^{-7}\) was reached on the SG points or the error ceased to decrease, a condition known as ``early stopping'' in the machine learning literature.
Table~\ref{tab:sg_performance_4d} presents Euler equation residuals for the four-dimensional (2-country) IRBC model, computed over a $10,000$-step simulated trajectory (discarding $1,000$ burn-in steps) from the stochastic steady state. As expected, both the maximum and average errors decrease consistently with higher SG refinement levels, remaining reasonably low even with modest grid point counts.
\begin{table}[t!]
    \centering
    \begin{tabular}{ccccc}
        \hline
        \textbf{\# Dimensions} & \textbf{SG Level} & \textbf{\# of Points} & \textbf{Avg. Error} & \textbf{Max. error (99.9\%)} \\
        \hline
         4 & 3 &  137 & -3.62 & -2.61 \\
         4 & 5 & 1105 & -4.21 & -3.10 \\
         4 & 7 & 7,537 & -4.64 & -3.31 \\
        \hline
    \end{tabular}
    \caption{Average (Avg.) and maximum (Max.) Euler equation residuals (i.e., errors) for the $4$-dimensional IRBC model across increasing SG refinement levels (SG Level), with corresponding grid point counts. All errors are reported in $\log_{10}$ scale.}
    \label{tab:sg_performance_4d}
\end{table}
We next increase the problem dimensionality from $d=4$ to 
$d=16$ while holding the grid level constant. 
Table~\ref{tab:sg_performance_dim} indicates that performance remains relatively consistent despite significant dimensional growth. Furthermore, the quality of the results reported here is at least on par with that achieved by other global solution methods (e.g., \citealp{juillard_et_al_2011}, and references therein).
\begin{table}[ht]
    \centering
    \begin{tabular}{cccccr}
        \hline
        \textbf{\# Dimensions} & \textbf{SG Level} & \textbf{\# of Points} & \textbf{Avg. Error} & \textbf{Max. error (99.9\%)} & \textbf{Time (s/step)} \\
        \hline
         4 & 3 & 137 & -3.62 & -2.61 & 0.13 \\
         8 & 3 & 849 & -3.78 & -2.71 & 2.09 \\
        16 & 3 & 6,049 & -4.05 & -2.94 & 267.35\\
        \hline
    \end{tabular}
\caption{Average (Avg.) and maximum (Max.) solution errors for $4$- to $16$-dimensional IRBC models, alongside the corresponding grid point counts for SGs at a fixed refinement level $3$, reported on a $\log_{10}$ scale, with indicative average runtimes in seconds per time iteration step ($\mathrm{s}/\mathrm{step}$).}
    \label{tab:sg_performance_dim}
    \footnotesize 
\end{table}
%
Table~\ref{tab:sg_performance_dim} shows that global solutions for $4$ and $8$-dimensional models require only seconds to few minutes, whereas $16$-dimensional cases demand about half an hour (cf. Table~\ref{tab:sg_performance_TI}).\footnote{Although a detailed discussion is beyond the scope of this paper, one could accelerate the time iteration algorithm by initially employing coarse SGs and progressively refining them to finer levels. For instance, \citet{brummetal_2017} show that SGs can reduce computation time by an order of magnitude by using a level-2 grid for 200 iterations, followed by 80 iterations on a level-3 grid or higher. 
}
Although the increase in runtime is notable, it exhibits subexponential growth and can, if needed, further be reduced by orders of magnitude through the use of high-performance computing resources~\citep{scheideggeretal_2018}.

Next, we demonstrate that a well-informed initial guess for the time iteration algorithm (Algorithm~\ref{alg:time_iteration}) significantly accelerates convergence (i.e., drastically reduces the time to solution). Table~\ref{tab:sg_performance_TI} shows that initializing the computations with Dynare’s linear solution reduces iteration steps by factors of approximately 5 (37 vs. 188 steps) in the $4$-dimensional model and nearly 20 times (5 vs. 94 steps) in the $16$-dimensional IRBC, compared to a naive, constant initial guess 
(\(\mathcal{I}p_{\text{guess}} = 1\) for all individual policies). Moreover, a good initial guess becomes increasingly important with rising dimensionality.
\begin{table}[ht]
    \centering
    \begin{tabular}{ccccc}
        \hline
        \textbf{\# Dimensions} & \textbf{SG Level} & \textbf{\# of Points} & \textbf{\# TI steps Dyn. Guess} & \textbf{\# TI Steps Naive Guess}\\
        \hline
         4 & 3 & 137 & 37 & 188 \\
         8 & 3 & 849 & 11 & 168\\
        16 & 3 & 6,049 & 5 & 94\\
        \hline
    \end{tabular}
\caption{The table reports the number of iteration steps required by Algorithm~\ref{alg:time_iteration} to achieve the target tolerance (\(\texttt{tol}\)), as evaluated on SG points at refinement level 3, for models of increasing dimensionality. The columns labeled ``\# TI Steps Dyn. Guess'' and ``\# TI Steps Naive Guess'' indicate the number of steps for models initialized with the linear Dynare solution and with a naive, constant initial guess, respectively.}
     \label{tab:sg_performance_TI}
\end{table}

A severe drawback of SGs of all types, including Smolyak’s method~\citep{kruegerkubler_2004}, is that the number of grid points grows very fast with the level of the approximation,
as was shown in Figure~\ref{fig:1}. It is, therefore, often not practical to increase accuracy by simply going to the next level. For instance, in 50 dimensions, a refinement level-3 SG comprises approximately 5,000 points, whereas a level-4 grid contains roughly 170,000 points, resulting in a substantial increase in computational burden. Adaptive sparse grids can resolve this problem to some extent, as intermediate grid sizes can be
reached by choosing the refinement threshold and maximum refinement level appropriately (cf.~\cite{brummscheidegger_2017}). However, for very high-dimensional problems (say, $d>20$), even they can fail. We, therefore, explore DDSG in Section~\ref{sec:HDMR_results}, which addresses this limitation by exploiting latent, additively separable structures in the model.

\subsection{Solving the IRBC model with DDSG and Dynare}
\label{sec:HDMR_results}

Having assessed the accuracy of our global solutions to the IRBC model and its scaling with SG resolution, we now systematically investigate how the performance of the DDSG method scales with increasing problem dimensionality. In accordance with our findings in Section~\ref{sec:SG_results} (see Table~\ref{tab:sg_performance_dim}), we fix the SG refinement level of each DDSG component function to $3$, since a level 3 SG previously yielded accurate results. Given the relative simplicity of the IRBC model under consideration, we conjecture that an additively separable structure is present and, accordingly, truncate the DDSG expansion at \(\ekk = 1\).

Table~\ref{tab:hdmr_performance} reports the Euler equation residuals for the IRBC model across dimensions ranging from four (2-country) to one hundred (50-country). These residuals are computed over a simulated trajectory of $10,000$ steps, with the first $1,000$ steps discarded as burn-in, starting from the stochastic steady state. Both the average and maximum errors remain consistently low across all dimensions, indicating robust performance. As discussed above, the quality of the results reported here is at least comparable to that achieved by other global solution methods for similar IRBC models, while operating on much larger state spaces (e.g., \citealp{juillard_et_al_2011} and references therein).

Furthermore, this table indicates that runtimes increase more gradually than in the pure SG case (see Table~\ref{tab:sg_performance_dim}), confirming that evaluating $N$ one-dimensional component functions is far more efficient for large $N$ than constructing an $N$-dimensional SG when a latent additively separable structure exists. For the four-dimensional model, the DDSG solution requires approximately forty percent more time per timestep due to the additional overhead required to carry around multiple grid structures in the computer memory. However, this overhead diminishes in the eight-dimensional case, where runtimes are roughly equivalent (i.e., slightly favoring DDSG). In the sixteen-dimensional case, a single DDSG timestep is over thirteen times faster. This advantage will become increasingly pronounced for higher dimensions, implying that DDSG keeps models tractable in scenarios where SG methods become infeasible.\footnote{Note that these figures are conservative, as our code remains unoptimized.}

%
%
\begin{table}[t]
    \centering
    \begin{tabular}{ccccr}
        \hline
        \textbf{\# Dimensions}  & \textbf{\# of Points} & \textbf{Avg. Error} & \textbf{Max. error (99.9\%)} & \textbf{Time (s/step)} \\
        \hline
         4 & 36 & -3.74 & -2.56 & 0.19 \\
         8 & 72 & -3.91 & -2.67 & 1.62 \\
         16 & 144 & -3.97 & -2.88 & 19.20 \\
        50 & 450 & -3.91 & -2.48 & 1849.41 \\
        100 & 900 & -3.97 & -2.45 & 7165.30 \\
        \hline
    \end{tabular}
    \caption{Average (Avg.) and maximum (Max.) errors for IRBC models of dimensions $4$ to $100$, alongside grid point counts for DDSG expansions with order $\ekk=1$, and the SG component functions fixed at a refinement level $3$. Errors and counts are reported on a $\log_{10}$ scale, with indicative average runtimes per time step in seconds ($\mathrm{s}/\mathrm{step}$).}
    \label{tab:hdmr_performance}
\end{table}
%

Next, we show that employing a well-informed initial guess for the time iteration algorithm (Algorithm~\ref{alg:time_iteration}) significantly accelerates convergence within the DDSG framework relative to a naive, constant initial guess (i.e., \(\mathcal{I}p_{\text{guess}} = 1\) for all individual policies), consistent with previous findings.
To ensure comparability between the SG and DDSG methods, we examine the four-, eight-, and sixteen-dimensional cases. 
\begin{table}[th]
    \centering
    \begin{tabular}{cccc}
        \hline
        \textbf{\# Dimensions} &  \textbf{\# of Points} & \textbf{\# TI steps Dyn. Guess} & \textbf{\# TI Steps Naive Guess}\\
        \hline
         4 & 36 & 7 & 209 \\
         8 & 72 & 3 & 186\\
        16 & 144 & 2 & 167\\
        \hline
    \end{tabular}
\caption{Number of iteration steps required by Algorithm~\ref{alg:time_iteration} to reach the target tolerance ($\texttt{tol}$), evaluated at DDSG points with $\ekk=1$, and SG refinement level 3, for IRBC models of increasing dimensionality. The columns ``{\# TI Steps Dyn. Guess}'' and ``{\# TI Steps Naive Guess}'' report the steps for models initialized with Dynare's linear solution and a naive zero guess, respectively.}
     \label{tab:ddsg_performance_TI}
\end{table}
Table~\ref{tab:ddsg_performance_TI} reveals that initializing computations with Dynare's linear solution reduces the number of iterations by factors of approximately 5 (7 vs. 209 steps) in the four-dimensional model and 80 (2 vs. 167 steps) in the sixteen-dimensional IRBC, relative to a naive, constant initial guess. Moreover, the importance of an effective initial guess grows with increasing dimensionality. As observed in the SG case, the initial guess becomes more impactful in higher dimensions; however, in the DDSG framework, Dynare's solution proves even more effective than in the SG case (see Table~\ref{tab:sg_performance_TI}).

\section{Conclusion \& Outlook} 
\label{sec:conclusion}

This study enhances Dynare (\texttt{Dynare.jl}) by integrating sparse grids and high-dimensional model representation into its native \texttt{.mod} files with minimal syntactic changes, enabling global solutions for high-dimensional, non-linear dynamic stochastic models. 
We implement these techniques in a generic manner, thereby substantially broadening the range of models that Dynare can tackle, and subsequently validate them using an international real business cycle model.
Our numerical experiments demonstrate their scalability, handling up to at least 100 dimensions on standard laptops within minutes to hours. 
Our methods mitigate the curse of dimensionality, reduce the time to solution by leveraging Dynare's perturbation-based initial guesses (yielding speedups of up to $80$ times), and maintain consistent accuracy as model dimensionality increases.



We conclude that these methodological additions significantly enhance Dynare's versatility, reinforcing its role as a pivotal tool for macroeconomic analysis. Looking forward, further advancements could be achieved by incorporating recent algorithms, such as those developed in the machine learning literature, thereby expanding Dynare's capacity to address increasingly intricate economic questions and preserving its continued relevance in the field.

\clearpage

\appendix

\clearpage
\bibliography{merged_references}{}
\bibliographystyle{plainnat}

	\end{singlespace}

\end{document}